\documentclass[intlimits,twoside,a4paper]{article}

\usepackage{amsmath,amssymb}
\usepackage{graphicx}

\usepackage[T2A]{fontenc}
\usepackage[cp1251]{inputenc}

\usepackage[eqsecnum]{cmpj2}

\issue{2014}{17}{1}{13602}
\doinumber{10.5488/CMP.17.13602}

\title[Interaction of phonons at superfluid helium-solid interfaces]%
{Interaction of phonons at superfluid helium-solid interfaces%
}

\author[I.N. Adamenko, E.K. Nemchenko]{I.N. Adamenko, E.K. Nemchenko}

\address {V.N. Karazin Kharkiv National University, 4 Svobody Sqr., 61022 Kharkiv, Ukraine}

\date{Received December 11, 2013}

\begin{document}

\maketitle

\begin{abstract}
A new method of obtaining the interaction Hamiltonian of phonons at superfluid helium-solid interface is proposed in the work.
Equations of hydrodynamic variables are obtained in terms of second quantization if helium occupies a half-space.
The contributions of all processes to the heat flux from solid to superfluid helium are calculated based on the obtained Hamiltonian.
The angular distribution of phonons  emitted by a solid is found in different processes. It is shown that all the exit angles
of superfuild helium phonons are allowed. The obtained results are compared with experimental data and with previous theoretical works.
\pacs 67.25.dt
\end{abstract}

\section{Introduction}

Superfluid helium has a whole number of unique phenomena that take place at superfluid helium-solid interface.
One of such phenomena is the thermal boundary resistance discovered by Kapitza~P.L.~\cite{Kap}.
It was discovered that there is a constant temperature difference between the contacting solid and superfluid helium
when a solid emits heat. Since then, this phenomenon has been studied by different authors because so far there
is no satisfactory agreement between experimental data and theoretical research.

The first theoretical explanation of Kapitza gap was given by Khalatnikov \cite{Khal2, Khal3, Khal4}.
 According to works \cite{Khal2, Khal3, Khal4}, the  heat flow occurs due to incident phonons in both superfluid
 helium and a solid.
These phonons with difficulty pass  through the interface due to acoustic mismatch of the media and due to the smallness of
incident phonon angle in liquid helium above which total internal reflection occurs. Transition probability
of phonon from one media to another which was obtained in \cite{Khal2, Khal3, Khal4} is proportional to
interface impedance $\rho_{\mathrm{L}}c_{\mathrm{L}} / \rho_{\mathrm{S}}c_{\mathrm{S}}$. Critical angle is equal
to $c_{\mathrm{L}} / c_{\mathrm{S}}$, where $c_{\mathrm{L}}$ and $c_{\mathrm{S}}$ are the velocities of sound of
liquid and solid, respectively, $\rho_{\mathrm{L}}$ and $\rho_{\mathrm{S}}$ are densities of liquid and solid, respectively.

The results of many experiments obtained by various authors significantly  differed from the calculated
values of theories \cite{Khal2, Khal3, Khal4}. Particularly, experimental values of heat transfer rate of
superfluid helium-solid interface are more often by two orders of magnitude larger than theoretical values in works \cite{Khal2, Khal3, Khal4}.

Large experimental values of heat transfer rate mean that there are other mechanisms of heat
transfer between superfluid helium and solid along with the so-called acoustic channel that was
considered in \cite{Khal2, Khal3, Khal4}. To our best knowledge, all theoretical works
dedicated to the the search for such mechanisms were based on the fact that the interface with superfluid
helium surface of solid was not perfectly smooth and clean and contained roughness, various defects and monolayers.

The imperfection of an interface leads to the assumption that phonons could pass into a solid at any incident
angles and not just in a narrow cone with a solid angle $\left ( c_{\mathrm{L}} / c_{\mathrm{S}} \right )^2$
which was formed by a critical angle following from Khalatnikov theory  \cite{Khal2, Khal3, Khal4}.
In this case, heat transfer rate may increase by $\left ( c_{\mathrm{S}} / c_{\mathrm{L}} \right )^2$ times.
This value is of the order of $10^2$ for the superfluid helium-solid interface. This fact can reconcile the
theory with experiments.

In this regard, numerous experiments were carried out by different authors in which the role of a solid surface
in heat transfer between solid and superfluid helium was investigated. The results of such experiments
are given in the review \cite{SwartzPohl} and in later experimental works \cite{Amrit6, Amrit7, Amrit8, Amrit9}.
These experiments indicated that the condition of a solid surface does substantially alter the heat transfer coefficient
so that it becomes larger than only an order of magnitude of the coefficient calculated in theory \cite{Khal2, Khal3, Khal4}.

\begin{figure}[htb]
\centerline{\includegraphics[width=0.65\textwidth]{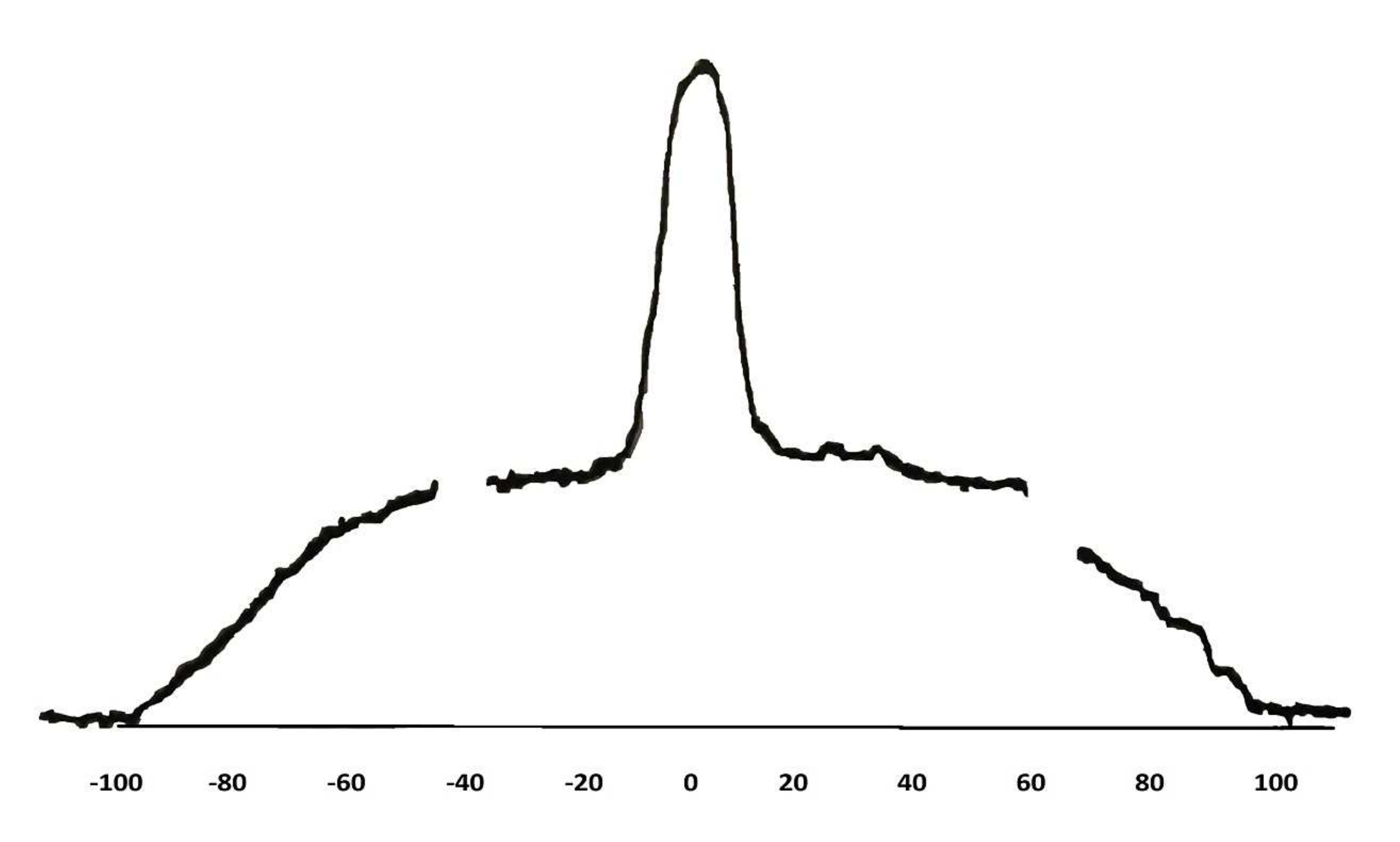}}
\caption{Angular distribution of heat flow from heated solid to superfluid that is observed in \cite{Wyatt10, Wyatt11, Wyatt12, Wyatt13, Wyatt14}.}
 \label{fig1}
\end{figure}

In order to understand the Kapitza gap problem, direct experiments
\cite{Wyatt10, Wyatt11, Wyatt12, Wyatt13, Wyatt14}
were performed in which energy and angular distribution of the emitted phonons by a solid in cold ($T < 100$~mK) superfluid
helium were measured. In these experiments, phonon beams were emitted from a heated solid to superfluid helium that
was almost at zero temperature (i.e., superfluid vacuum). As the heaters there were used both conductive metal films and cleaved
surfaces of crystals that were almost perfect surfaces. It was shown in works  \cite{Wyatt10, Wyatt11, Wyatt12, Wyatt13, Wyatt14}
 that even with almost perfect solid surface there are two channels of phonon transfer from solid to superfluid
 helium which are demonstrated in figure~\ref{fig1}

The first channel formed a sharp peak of phonons emitted in a narrow cone of angles whose axis was normal
to the solid surface (see figure~\ref{fig1}). The observed value in \cite{Wyatt11} of the angle of the cone
coincides with the calculated values for different solids in the acoustic mismatch theory which was based on Khalatnikov
theory \cite{Khal2, Khal3, Khal4}. This channel was called the acoustic channel.

The second channel, the so-called background channel, contained phonons emitted in all
directions. Moreover, it was shown that the contribution of a background channel was an order of
magnitude larger than the contribution of an acoustic channel during experimental data analysis.

In work \cite{Adam15} which was performed based on the results of experimental work \cite{Wyatt16}
it was shown that the phonons emitted by a heated and rather rough gold surface in superfluid helium were
also distributed through two channels observed in  \cite{Wyatt10, Wyatt11, Wyatt12, Wyatt13, Wyatt14}.

Accordingly, a question has arisen: what is the physical reason for the existence of such a
large background channel at almost perfect surface solid? Great hope to explain the existence of the
background channel and large observed values of the heat transfer coefficient in Kapitza gap experiments
was entrusted to the processes in which there was a different number of phonons in the initial and the final states.
These are the so-called inelastic interaction processes. A possible diagram of inelastic process can be found in
the experimental work \cite{Wyatt12}, where one phonon of a solid transforms into two phonons of a liquid that
could pass at any angles to the interface. One of the possible inelastic processes that differs from the one illustrated
in \cite{Wyatt12} was considered by Khalatnikov \cite{Khal4} who showed that the contribution of this process
was relatively small. It is worth pointing out that the inelastic process considered in \cite{Khal4} does not
contribute to the heat flux from solid to superfluid helium which is almost at zero temperature.

In this regard, consideration of all possible inelastic processes turns out to be relevant as well as the calculation of their
contribution to the background channel. This is the focus of the present work.

The first attempts to solve the above mentioned problem was made in works \cite{ShBowley, Zhukov} in which it was suggested
to create a microscopical theory of Kapitza gap at the He II-solid interface. However, it was a failure to create
a self-consistent approach capable of  yielding the results in accord with the acoustic theory \cite{Khal2, Khal3, Khal4}
corresponding to elastic phonon processes. This is apparently connected with the calculations
that were not brought to final analytical formulas and to specific numerical values in works \cite{ShBowley} and \cite{Zhukov}.

The original results of constructing a unified self-consistent theory describing
both elastic and inelastic processes at the superfluid helium-solid interface were presented
at the QFS2012 conference\footnote{QFS2012: International Conference on Quantum Fluids and Solids,
15--21 August 2012, Physics Department, Lancaster University, UK.} by the authors
of this paper. These first results were published in the materials of the conference \cite{jltp_prev}.
The contribution of inelastic processes to Kapitza gap was considered in the work \cite{ilt_prev}.

The main goal of this paper is to investigate all possible inelastic processes that contribute to
the heat flow from the solid to the superfluid helium and to consider the angular distribution
of the emitted phonons in different processes.

\section{Interaction Hamiltonian of helium phonons with an oscillating surface of a solid}

For the interaction Hamiltonian of helium phonons with an oscillating surface of a solid, we calculate the density of
the energy of superfluid helium in the presence of an oscillating interface. The obtained Hamiltonian will essentially
differ from the Hamiltonians  used in works \cite {Khal4, ShBowley, Zhukov} and will yield a correct result
regarding the heat flow due to the elastic process that is equal to the result obtained in \cite {Khal3}.

Oscillations of interface excite in helium oscillations of density $\rho_{\mathrm{i}}$ and velocity $\mathbf{v}_{\mathrm{i}}$
along with the intrinsic oscillations of $\rho$ and $\mathbf{v}$ in the liquid. In this case, the interaction energy is
\begin{equation}
\label{eq1}
E = \frac{1}{2}
	\left ( \rho_{\mathrm{L}} + \rho + \rho_{\mathrm{i}} \right )
	\left ( \mathbf{v} + \mathbf{v}_{\mathrm{i}} \right )^2
	+
	E_\rho
	\left (\rho_{\mathrm{L}} + \rho + \rho_{\mathrm{i}} \right  ),
\end{equation}
where $E_\rho$ is the density functional.

To simplify the problem, we restrict ourselves to longitudinal phonons in the solid. The inclusion of
transverse phonons does not cause fundamental difficulties, but all the calculations become more
cumbersome and lead to the appearance of a factor $F$ in the final calculations that depends on
the elastic constants of the solid. $F$ varies over small limits and remains of the order of $2$ for different solids.

Now we reduce the equation~\eqref{eq1} to the form of an expansion accurate to cubic terms in the small
parameters $\rho_{\mathrm{i}}$, $\mathbf{v}_{\mathrm{i}}$, $\rho$ and $\mathbf{v}$:
\begin{equation}
\label{eq2}
E = E_{0,1} +
	\frac{1}{2}\rho_{\mathrm{L}}\left ( \mathbf{v} + \mathbf{v}_{\mathrm{i}} \right )^2 +
	\frac{c_{\mathrm{L}}^2}{2\rho_{\mathrm{L}}}\left( \rho + \rho_{\mathrm{i}}\right) ^2 +
	\frac{1}{2}\left( \rho + \rho_{\mathrm{i}}\right) \left( \mathbf{v} + \mathbf{v}_{\mathrm{i}} \right )^2  +
	\frac{c_{\mathrm{L}}^2}{6\rho_{\mathrm{L}}^2} \left( 2u-1\right) \left( \rho + \rho_{\mathrm{i}}\right)^3\,,
\end{equation}
where $u=
	\frac{\rho_{\mathrm{L}}}{c_{\mathrm{L}}}
	\frac{\partial c_{\mathrm{L}}}{\partial \rho_{\mathrm{L}}}$
is the Gruneisen constant, which equals 2.84 for helium,
\begin{equation}
\label{eq3}
E_{0,1} = E_\rho \left ( \rho_{\mathrm{L}} \right) +
	\Bigl. \frac{\partial E_\rho \left( \rho_{\mathrm{t}} \right)} {\partial \rho_{\mathrm{t}}} \Bigr|_{\rho_{\mathrm{t}}=\rho_{\mathrm{L}}}
	\left( \rho + \rho_{\mathrm{i}} \right)
\end{equation}
is the sum of zero and the first terms of the expansion which does not contribute to the interaction of
the liquid and the solid, $\rho_{\mathrm{t}} = \rho_{\mathrm{L}} + \rho + \rho_{\mathrm{i}}$.

Then, the contribution to the interaction of helium with a wall will yield a term that
simultaneously contains parameters characterizing both the solid and the liquid. In this case, the interaction energy is as follows:
\begin{equation}
\label{eq4}
E_{\mathrm{int}} = \rho_{\mathrm{L}}\mathbf{v}\mathbf{v}_{\mathrm{i}} +
	\frac{c_{\mathrm{L}}^2}{\rho_{\mathrm{L}}}\rho\rho_{\mathrm{i}} +
	\frac{\rho}{2} \left( 2\mathbf{v}\mathbf{v}_{\mathrm{i}} + \mathbf{v}_{\mathrm{i}}^2 \right) +
	\frac{\rho_{\mathrm{i}}}{2} \left( 2\mathbf{v}\mathbf{v}_{\mathrm{i}} + \mathbf{v}^2 \right) +
	\frac{c_{\mathrm{L}}^2}{2\rho_{\mathrm{L}}^2}\left( 2u-1 \right) \rho\rho_{\mathrm{i}} \left( \rho + \rho_{\mathrm{i}} \right).
\end{equation}

The first two terms in the equation~\eqref{eq4} describe the two-phonon interactions and the
remaining terms describe the three-phonon interactions (in terms of secondary quantization).

In this problem, the velocity and density of solid and liquid phonons are specified
for a half space,
whereas there are problems expanding them in Fourier series and with the subsequent use of the second quantization method.
The following method for analytic continuation of the solutions is proposed to overcome these difficulties
and make it possible to use the Fourier expansion and secondary quantization. To this end, we carry out
calculations on the entire axis $z$ that is perpendicular to the superfluid helium-solid interface.
Moreover, due to boundary conditions at $z=0$, $v_z$ is oddly extended to the entire space so that
	$v_z\left(z > 0\right) = -v_z\left(z < 0\right)$
and
	$v_x$, $v_y$, $\rho$ and $v_{{\mathrm{i}}z}$
are evenly extended to the entire space.

However, we should note that the helium perturbations generated by oscillations of the interface,
on the one hand, contribute to the energy of helium, and, on the other hand, are determined by the
parameters which characterize the vibrations of a solid interface (amplitude and displacement velocity).
The relationship between these parameters is given by standard boundary conditions for a normal component of
the velocity at the solid-superfluid liquid interface, which is superfluid helium. Thus, parameters
describing the vibrations of the interface after the second quantization and the change of helium energy
caused by these vibrations  will contain  the creation and annihilation operators of solid phonons.
In this respect, those perturbation operators of density and velocity of helium and velocity of interface
vibrations are Hermitian after the second quantization. We get the final form of these operators:
\begin{align}
\label{eq5}
\hat{\rho} &= \rho_{\mathrm{L}}\sum^{+\infty}_{k_z=0} \sum_{\mathbf{k}_{||}}	
{
		\frac{\ri}{c_{\mathrm{L}}}
		\sqrt {\frac{\hbar\omega} {2\rho_{\mathrm{L}}V_{\mathrm{L}}}}
		\Bigl(
			\hat{a}_\mathbf{k} - \hat{a}_\mathbf{-k}^+ + \hat{a}_\mathbf{-k} - \hat{a}_\mathbf{k}^+
		\Bigr)	
		\left(
			\frac{\re^{\ri k_zz} + \re^{-\ri k_zz}} {\sqrt{2}}
		\right)
		\re^{\ri  \mathbf{k}_{||} \mathbf{r}_{||}}
},
\nonumber\\
\hat{v}_z &=\sum^{+\infty}_{k_z=0} \sum_{\mathbf{k}_{||}}	
{
		\sqrt {\frac{\hbar\omega} {2\rho_{\mathrm{L}}V_{\mathrm{L}}}}
		\ri\frac{k_z}{k}
		\Bigl(
			\hat{a}_\mathbf{k} + \hat{a}_\mathbf{-k}^+ + \hat{a}_\mathbf{-k} + \hat{a}_\mathbf{k}^+
		\Bigr)	
		\left(
			\frac{\re^{\ri k_zz} - \re^{-\ri k_zz}} {\sqrt{2}}
		\right)
		\re^{\ri  \mathbf{k}_{||} \mathbf{r}_{||}}
},
\nonumber\\
\hat{v}_{iz} &=\sum^{+\infty}_{q_z=0} \sum_{\mathbf{q}_{||}}	
{
		\sqrt {\frac{\hbar\Omega} {2\rho_{\mathrm{S}}V_{\mathrm{S}}}}
		\ri\frac{q_z}{q}
		\Bigl(
			\hat{b}_\mathbf{q} - \hat{b}_\mathbf{-q}^+ + \hat{b}_\mathbf{-q} - \hat{b}_\mathbf{q}^+
		\Bigr)	
		\left(
			\frac{\re^{\ri b_zz} + \re^{-\ri b_zz}} {\sqrt{2}}
		\right)
		\re^{\ri  \mathbf{q}_{||} \mathbf{r}_{||}}
},
\end{align}
where $\mathbf{k}$ and $\mathbf{q}$ are wave vectors of helium and solid phonons, respectively,
$\omega$ and $\Omega$ are frequencies of helium and solid phonons, respectively,
$V_{\mathrm{L}}$ and $V_{\mathrm{S}}$ are volumes that liquid and solid occupies,
$\hat{a}_\mathbf{k}^+\left(  \hat{a}_\mathbf{k} \right)$  and $\hat{b}_\mathbf{q}^+\left(  \hat{b}_\mathbf{q} \right)$
are operators of creation (annihilation) of helium and solid phonons, respectively; axis $z$ is directed perpendicular to the interface,
and $\mathbf{k}_{||}$ and $\mathbf{q}_{||}$ tangential components of the wave vectors of helium and solid phonons, respectively.
Equations ~\eqref{eq4} and \eqref{eq5} permit to submit Hamilton operator
\begin{equation}
\label{eq6}
\hat{H}_{\mathrm{int}} =
	\int\limits_0^L \rd z \int \rd S E_{\mathrm{int}}
\end{equation}
in terms of the second quantization. In equation~\eqref{eq6}, integration is over the volume
of the liquid $V_{\mathrm{L}}=LS$, where $S$ is the area of the superfluid helium-solid interface.
The Hamiltonian equation~\eqref{eq6} describes the creation and annihilation of phonons at the He II-solid interface,
 which is caused by vibrations of the interface.

After these procedures, the Hamiltonian \eqref{eq6} will have the following form to within the cubic terms
\begin{equation}
\label{eq7}
\hat{H}_{\mathrm{int}} =
	\hat{H}^{(2)}_{\mathrm{int}} + \hat{H}^{(3)}_{\mathrm{int}}\,.
\end{equation}

Here, the first term
\begin{equation}
\label{eq8}
\hat{H}^{(2)}_{\mathrm{int}} =
	\ri c_{\mathrm{L}}\sqrt{\frac{\rho_{\mathrm{L}}}{\rho_{\mathrm{S}}}}\frac{\hbar S}{\sqrt{V_{\mathrm{L}}V_{\mathrm{S}}}}
	\sum_{\mathbf{k}}\sum_{\mathbf{q}}
	{
		\frac{q_z}{q}
		\Bigl(
			\hat{a}_\mathbf{k} + \hat{a}_\mathbf{-k}^+ + \hat{a}_\mathbf{-k} + \hat{a}_\mathbf{k}^+
		\Bigr)	
		\Bigl(
			\hat{b}_\mathbf{q} - \hat{b}_\mathbf{-q}^+ + \hat{b}_\mathbf{-q} - \hat{b}_\mathbf{q}^+
		\Bigr)	
		\delta_{\mathbf{k}_{||},\mathbf{q}_{||}}
	}
\end{equation}
contains a single phonon annihilation (creation) operator and a single creation (annihilation) operator for the solid.

Thus, $\hat{H}^{(2)}_{\mathrm{int}}$ describes the conversion of a liquid (solid) phonon into
a solid (liquid) phonon at the superfluid helium-solid interface. In this transition, the
phonon retains its energy. We refer to this kind of a process as an elastic one. The second term in equation~\eqref{eq7} has the form
\begin{eqnarray}
\label{eq9}
\lefteqn{\hat{H}^{(3)}_{\mathrm{int}} =
	\frac{\hbar^{3/2}S}{c_{\mathrm{L}}V_{\mathrm{L}}\sqrt{V_{\mathrm{S}}\rho_{\mathrm{S}}}} \sum_{\mathbf{k},\mathbf{q}}
	{
		\sqrt{\omega_1\omega_2\Omega}\frac{k_{2z}}{k_2}\frac{q_z}{q}
		\frac{k_{2z}}{k_{2z}^2 - k_{1z}^2}
		\delta_{\mathbf{k}_{1||}+\mathbf{k}_{2||}+\mathbf{q}_{||},0}
		\Bigl(
			\hat{a}_{\mathbf{k}_1} - \hat{a}_{\mathbf{-k}_1}^+ + \hat{a}_{\mathbf{-k}_1} - \hat{a}_{\mathbf{k}_1}^+
		\Bigr)	
	}}
\nonumber\\
&&{}\times
\Bigl(
			\hat{a}_{\mathbf{k}_2} + \hat{a}_{\mathbf{-k}_2}^+ + \hat{a}_{\mathbf{-k}_2}
			+ \hat{a}_{\mathbf{k}_2}^+
		\Bigr)	
		\Bigl(
			\hat{b}_\mathbf{q} - \hat{b}_\mathbf{-q}^+ + \hat{b}_\mathbf{-q} - \hat{b}_\mathbf{q}^+
		\Bigr)	+
		\frac{\hbar^{3/2}\sqrt{\rho_{\mathrm{L}}}S}{c_{\mathrm{L}}V_{\mathrm{L}}\sqrt{V_{\mathrm{S}}}\rho_{\mathrm{S}}}
        \sum_{\mathbf{k},\mathbf{q}}
		\sqrt{\omega\Omega_1\Omega_2}\frac{1}{k}
		\frac{q_{1z}}{q_1}\frac{q_{2z}}{q_2}
		\delta_{\mathbf{q}_{1||}+\mathbf{q}_{2||}+\mathbf{k}_{||},0}
\nonumber\\
&&{}\times
\Bigl(
			\hat{a}_\mathbf{k} + \hat{a}_\mathbf{-k}^+ + \hat{a}_\mathbf{-k} + \hat{a}_\mathbf{k}^+
		\Bigr)	
		\Bigl(
			\hat{b}_{\mathbf{q}_1} - \hat{b}_{\mathbf{-q}_1}^+ + \hat{b}_{\mathbf{-q}_1} - \hat{b}_{\mathbf{q}_1}^+
		\Bigr)	
		\Bigl(
			\hat{b}_{\mathbf{q}_2} - \hat{b}_{\mathbf{-q}_2}^+ + \hat{b}_{\mathbf{-q}_2} - \hat{b}_{\mathbf{q}_2}^+
		\Bigr),
\end{eqnarray}
where $\omega_{1,2}$ and $\mathbf{k}_{1,2}$ are frequencies and wave vectors of the superfluid
helium phonons and $\Omega_{1,2}$ and $\mathbf{q}_{1,2}$   are frequencies and wave vectors of the solid.
Equation~\eqref{eq9} describes the processes in which there are different numbers of phonons in the initial
and final states. We shall refer to these kinds of processes as inelastic phonon processes.

\section{Heat flow through the superfluid helium-solid interface}

In order to calculate the heat flow from a solid to a liquid, it is necessary to write down the probability
of a phonon conversion process at the liquid helium-solid interface. From the Hamiltonian
equations~\eqref{eq7}, \eqref{eq8} and \eqref{eq9}, there are four possible three-phonon inelastic
processes along with one elastic process. We enumerate these processes with a subscript $k$ equal to $0$ for the
elastic process and $1\div4$ for the four possible inelastic processes. Here are diagrams of all possible processes.
The diagrams for reverse processes are obtained by reversing the directions of all the arrows in the diagram for a forward process.
\begin{figure}[htb]
\centerline{\includegraphics[width=0.4\textwidth]{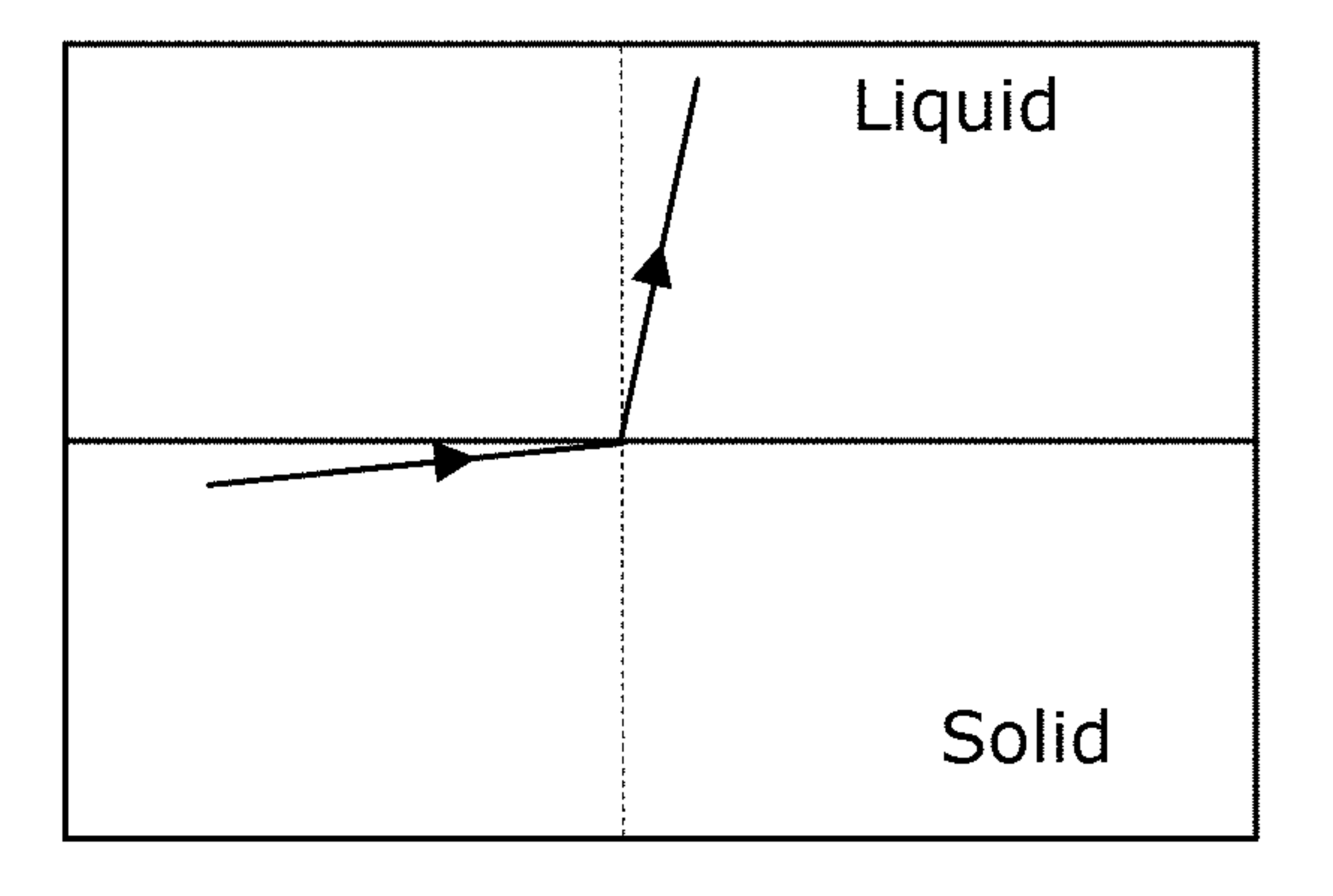}}
\caption{Diagram for a direct elastic phonon conversion process at a superfluid helium-solid interface $(k=0)$.}
\label{fig2}
\end{figure}
\begin{figure}[htb]
\centerline{\includegraphics[width=0.65\textwidth]{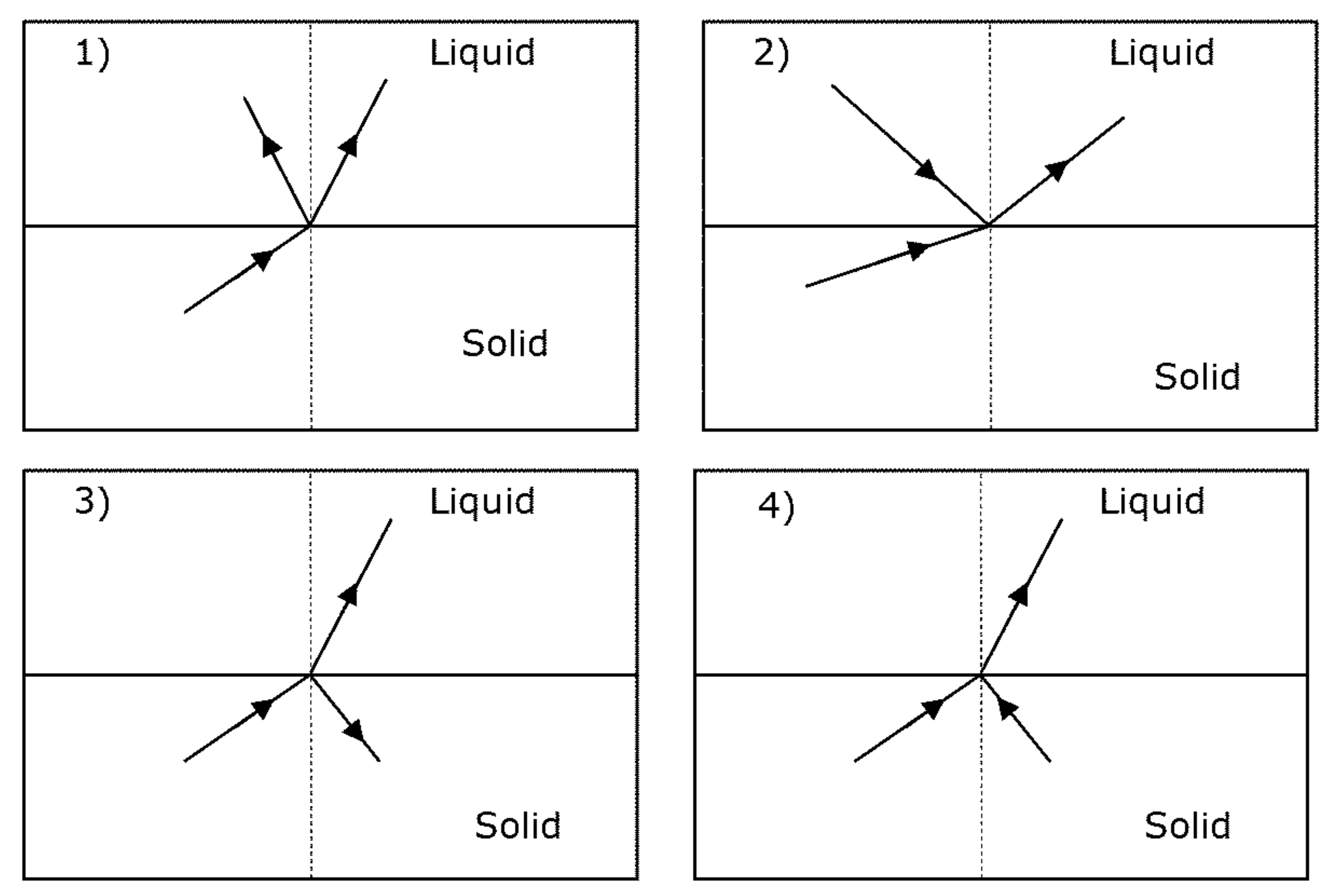}}
\caption{Diagrams for  direct inelastic processes $(k = 1 \div 4)$.}
 \label{fig3}
\end{figure}

The second inelastic process does not give a contribution to the heat flow from the heated solid to
superfluid helium that is at zero temperature. Therefore, we will consider only the first,
the third and the fourth inelastic processes along with the elastic process.

The probability $w_k$ of process $k$, which is determined by the matrix element
	$M_{{fi}}^{(k)} = \langle f \vert \hat{H}_{\mathrm{int}} \vert i \rangle$
for a transition from the initial state $i$ to the final state $f$, if a particular process
results from the Hamiltonian equation~\eqref{eq7}, is given by
\begin{equation}
\label{eq10}
	w_k=
		\frac{2\pi}{\hbar S}
		\Bigl|
			M_{{fi}}^{(k)}	
		\Bigr| ^2
		\delta
		\left(
			E_{{f}} - E_{{i}}
		\right).
\end{equation}
Here, $E_{{f}}$ and $E_{{i}}$ are the total energy of the phonons in the final
and initial states, respectively. The quantity~\eqref{eq10} is the probability that phonons transfer
from state $i$ into state $f$ per unit time through unit area of the interface surface.

The expression for a heat flux per unit time through unit area of the interface surface in the
normalization of the operators to the energy of a single phonon for the $k$-th process that we have chosen is
\begin{equation}
\label{eq11}
	W^{(k)}=
		\int w_k \sum_{{f}} \varepsilon_{{f}}
		\cos \theta_{{f}}
		\prod_{{f}}
		\left[
			1+n\left( \varepsilon_{{f}} \right)
		\right]
		\rd \Gamma_{{f}}
		\prod_{{i}}
		n\left( \varepsilon_{{i}} \right)
		\rd \Gamma_{{i}}\,,
\end{equation}
where the sum is taken over all the final phonons,
while the products $\prod_{{f}}$ and $\prod_{{i}}$ are taken over all the final
and initial phonons, respectively, $\varepsilon_{{f}}$ and $\varepsilon_{{i}}$ are
the energies of the final and initial phonons, respectively, $n(\varepsilon)$ is the Bose distribution function,
$\rd \Gamma = \rd^3p\rd^3r/ \left( 2\pi\hbar\right)^3$  is the number of quantum states in an
element of phase space, and $\theta_{{f}}$ is the exit angle for a final phonon with energy $\varepsilon_{{f}}$.
Here, and in what follows, all the angles are reckoned from the normal to the superfluid helium-solid interface boundary.

We consider the heat flow due to an elastic process. The matrix element of this process is as follows:
\begin{equation}
\label{eq12}
	M^{(0)}_{{fi}}=
		\frac {2\ri c_{\mathrm{L}}\hbar S} {\sqrt{V_{\mathrm{L}}V_{\mathrm{S}}}}
		\sqrt{\frac {\rho_{\mathrm{L}}} {\rho_{\mathrm{S}}}} \frac {q_z}{q}
		\delta_{\mathbf{k}_{||},\mathbf{q}_{||}}	\,.
\end{equation}

For the heat flow from a solid at temperature $T_{\mathrm{S}}$  into superfluid helium,
which is at zero temperature, we begin with equations~\eqref{eq10}, \eqref{eq11}, and \eqref{eq12} and obtain
\begin{equation}
\label{eq13}
	W^{(0)} =
		\frac{4\pi^4}{15}
		\frac{\rho_{\mathrm{L}}c_{\mathrm{L}}}{\rho_{\mathrm{S}}c_{\mathrm{S}}^3}
		\frac{1}{3(2\pi)^2\hbar^3}
		\left(
			k_\mathrm{B}T_{\mathrm{S}}
		\right)^4\,.
\end{equation}

According to conservation of energy and conservation of tangential impulse component of phonon,
it follows that in an elastic process, the heat flux~\eqref{eq13} will fill a narrow cone of angles with solid angle
$\left(c_{\mathrm{L}} / c_{\mathrm{S}} \right)^2$,
whose axis is directed normal to the interface.

The first inelastic process, which corresponds to a transition from a state with one solid phonon
to a state with two liquid phonons, is calculated in a standard way and is as follows:
\begin{equation}
\label{eq14}
	M_{{fi}}^{(1)} =
		\frac {2\sqrt{2} \hbar^{\frac{3}{2}} S} {c_{\mathrm{L}} V_{\mathrm{L}} \sqrt {V_{\mathrm{S}} \rho_{\mathrm{S}}}}
		\sqrt{\omega_1 \omega_2 \Omega}
		\left[
			\frac {k^2_{2z}} {k_2 \left( k^2_{2z} - k^2_{1z}\right)} -
			\frac {k^2_{1z}} {k_1 \left( k^2_{1z} - k^2_{2z}\right)}
		\right]
		\frac {q_z}{q}
		\delta_{\mathbf{k}_{1||}+\mathbf{k}_{2||}+\mathbf{q}_{||},0}\,.
\end{equation}

On the assumption of~\eqref{eq10}, \eqref{eq11} and \eqref{eq14}, the heat flow from a solid at
temperature $T_{\mathrm{S}}$ into a liquid helium at zero temperature is as follows:
\begin{eqnarray}
\label{eq15}
	W^{(1)}(S\rightarrow L) &=&
		\frac {8}
			{(2\pi)^4 \rho_s c_{\mathrm{L}}^4 c_{\mathrm{S}}^3 \hbar^6}
		\left(
			k_b T_{\mathrm{S}}
		\right)^8
		\int
		\rd x \rd y \sin \theta \rd \theta \sin \theta_1 \rd \theta_1
		\cos^2 \theta  \frac {1} {\re^x - 1}
		\left[
			y \frac {\cos \theta_1} {\cos \theta_2} +
			(x - y)
		\right]
\nonumber\\
&&{}
		\times
		y^3 x^3 (x-y)
		\left[
			\frac
				{(x-y) \cos^2\theta_2 + y \cos^2\theta_1}
				{(x-y)^2\cos^2\theta_2 - y^2\cos^2\theta_1}
		\right]^2\, ,
\end{eqnarray}
where $x = \hbar\Omega / k_{\mathrm{B}} T_{\mathrm{S}}$,
$y = \hbar\omega_1 / k_{\mathrm{B}} T_{\mathrm{S}}$ and
$\hbar\omega_2 / k_{\mathrm{B}} T_{\mathrm{S}} = x-y$ are from the conservation of energy law,
$\theta_{1,2}$ are the exit angles of liquid phonons, $\theta$ is the incident angle of a solid phonon.
The numerical value of the dimensionless integral~\eqref{eq15}  is independent of temperature, but it
does depend on the ratio of the speeds of sound of the solid and the liquid. For the value $c_{\mathrm{L}} / c_{\mathrm{S}} = 0.1$,
that will be used further, dimensionless integral \eqref{eq15} is equal to $4.78\cdot 10^2$. As will be shown below,
this value weakly depends on the ratio $c_{\mathrm{L}} / c_{\mathrm{S}}$.

It should be noted that in integrals \eqref{eq15} and further, integration of the azimuthal angle
is replaced by multiplication by $2\pi$ for simplicity. The limits of integration in these integrals and the function
of the integration variables
$\theta_2 = \theta_2(\theta, \theta_1, x, y)$
are determined by the conservation of energy and by tangential component of impulse laws. Equation for $\sin\theta_2$ is as follows:
\begin{equation}
\label{eq16}
\sin\theta_2 =
	\frac {1} {x-y}
	\left(
		y\sin\theta_1 + x \frac {c_{\mathrm{L}}} {c_{\mathrm{S}}} \sin\theta
	\right).
\end{equation}

The conditions of exit phonon angles with energies $\omega_1$ could be obtained from equation~\eqref{eq16}.
\begin{enumerate}
\item$0\leqslant\theta_1\leqslant\frac {\pi} {2}$, for $x \leqslant \frac {y} {2}
\left( 1 - \frac{c_{\mathrm{L}}}{c_{\mathrm{S}}} \sin\theta\right).$
\item$0\leqslant\theta_1\leqslant \arcsin \left[ \frac{y}{x} \left( 1 - \frac{c_{\mathrm{L}}}{c_{\mathrm{S}}} \sin\theta\right) - 1\right]$,
 for $x \geqslant \frac {y} {2} \left( 1 - \frac{c_{\mathrm{L}}}{c_{\mathrm{S}}} \sin\theta\right).$
\end{enumerate}

Consider the limiting cases of these conditions.
\begin{itemize}
\item[a)]The elastic case: $\omega_1 = \Omega$, $\omega_2 = 0$. The condition on the exit angle is as follows:
\begin{equation}
\label{eq17}
0 \leqslant \theta_1 \leqslant \arcsin
\left(
	\frac {c_{\mathrm{L}}} {c_{\mathrm{S}}} \sin\theta	
\right),
\end{equation}
which coincides with the condition in the elastic process.
\item[b)]Weak-inelastic case: $\omega_1 = \Omega - \Delta$, $\omega_2 = \Delta$, considering that
$\Delta \ll  \Omega$, but $\Delta > \Omega \frac {c_{\mathrm{L}}}{c_{\mathrm{S}}} \sin \theta $.
The condition takes in account the smallness of $c_{\mathrm{L}} / c_{\mathrm{S}}$
\begin{equation}
\label{eq18}
0 \leqslant \theta_1 \leqslant \frac {\Delta}{\Omega}\,.
\end{equation}
\item[c)]Inelastic case: $\omega_1 = \omega_2 = \Omega$. The condition is as follows:
\begin{equation}
\label{eq19}
0 \leqslant \theta_1 \leqslant \frac{\pi}{2}\,.
\end{equation}
\end{itemize}

This shows that if the energies of the created phonons are equal to each other, all the exit phonon angles are
allowed in the first inelastic process. The ban on these angles is determined by the proximity of the liquid
phonon energy to the energy of a solid phonon. According to the conditions on the incident phonon angles,
the ratio of velocities $c_{\mathrm{L}}/c_{\mathrm{S}}$ gives a small contribution both to integration
limits and to the value of integral~\eqref{eq15}.

Unlike the elastic process, the phonons which were born in this inelastic process will move in all
directions relative to the normal to the interface. Then, the phonons that were emitted in all directions
should be observed in the angular distribution of phonons emitted by a heated solid to superfluid helium
along with a sharp acoustic peak (see figure~\ref{fig1}). The rate of the heat flow due to the elastic~\eqref{eq12} process
and the first inelastic~\eqref{eq15} process is as follows:

\begin{equation}
\label{eq20}
\frac {W^{(0)}}{W^{(1)}} =
	\frac {1} {1.08 \cdot 10^4}
	\frac {\pi^6 \rho_{\mathrm{L}} c^5_{\mathrm{L}} \hbar^3} {\left( k_{\mathrm{B}} T_{\mathrm{S}}\right)^4}\,.
\end{equation}

Equation~\eqref{eq20} shows that for $T_{\mathrm{S}}=5$~K, the heat flux through the superfluid helium-solid interface
produced by the first inelastic process is $2.3$ times greater than that produced by the elastic process.
This value is by a factor of four smaller than the one observed experimentally \cite{Wyatt10, Wyatt11, Wyatt12, Wyatt13, Wyatt14}.
For $T_{\mathrm{S}}=1$~K, the contribution of the heat flux from the first inelastic process is by a factor of $272$ less
than that from the elastic process. Thus, the first inelastic process cannot completely explain the relatively large
experimentally observed  \cite{Wyatt10, Wyatt11, Wyatt12, Wyatt13, Wyatt14} level of background emission.

Analogously, for the third process and the fourth process, the heat flow will be as follows:
\begin{align}
\label{eq21}
W^{(3)} &=
	\frac {32 \rho_{\mathrm{L}}} { (2\pi)^4 \rho^2_{\mathrm{S}} c^4_{\mathrm{S}} c^3_{\mathrm{L}} \hbar^6} \left( k_{\mathrm{B}} T_{\mathrm{S}}\right)^8
	\int
	\rd x \rd y \sin\theta \rd \theta \sin \theta_1 \rd \theta_1
	\frac {1} {\re^x - 1} y^2 x^3 (x-1) \cos \theta \cos \theta_2 \cos^2 \theta_1\,,
\\
\label{eq22}
W^{(4)} &=
	\frac {32 \rho_{\mathrm{L}}} { (2\pi)^4 \rho^2_{\mathrm{S}} c^4_{\mathrm{S}} c^3_{\mathrm{L}} \hbar^6 } \left( k_{\mathrm{B}} T_{\mathrm{S}}\right)^8
	\int
	\rd x \rd y \sin\theta \rd \theta \sin \theta_1 \rd \theta_1
	\frac {1} {\re^x - 1} \frac {1} {\re^{y-x} - 1}
	 y^2 x^3 (x-1) \cos \theta \cos \theta_2 \cos^2 \theta_1\,,\nonumber\\
\end{align}
where $x=\hbar\Omega_1/k_{\mathrm{B}} T_{\mathrm{S}}$,
$y=\hbar\omega/k_{\mathrm{B}} T_{\mathrm{S}}$ and $\hbar\Omega_2/k_{\mathrm{B}} T_{\mathrm{S}} = x-y$
from the conservation of energy law. Numerical values of integrals that are in \eqref{eq21} and \eqref{eq22}
are $6.32\cdot 10^3$ and $8.53\cdot 10^3$, respectively. The ratio of contributions of the third and the second
processes to the contribution of the first process due to \eqref{eq15}, \eqref{eq21} and \eqref{eq22} are, respectively, as follows:
\begin{eqnarray}
\label{eq23}
\frac {W^{(3)}} {W^{(1)}} =
	5.21 \frac {\rho_{\mathrm{L}} c_{\mathrm{L}}}{\rho_{\mathrm{S}} c_{\mathrm{S}}}\,, \qquad
\frac {W^{(4)}} {W^{(1)}} =
	7.07 \frac {\rho_{\mathrm{L}} c_{\mathrm{L}}}{\rho_{\mathrm{S}} c_{\mathrm{S}}}\,.
\end{eqnarray}

The investigation of angular phonon distribution emitted by a heated solid in the third and the fourth
processes similar to those that were presented for the first process shows that phonons are emitted
in all directions to the superfluid helium. According to equation \eqref{eq23}, the third and the fourth processes
give contributes into the heat flow from solid to superfluid helium of the same order of magnitude.
This contribution is by an order of magnitude less than contribution of the first process.

\section{Conclusion}

In this paper we have derived the interaction Hamiltonian \eqref{eq7}--\eqref{eq9} of phonons of superfluid
helium with an oscillating solid interface. The phonon field has been quantized in the half space, which made
it possible to write down this Hamiltonian in terms of annihilation and creation operators for phonons of the
superfluid helium and of the solid.

The probabilities both of the elastic process and all of the inelastic processes have been calculated from the Hamiltonian.
The derived equations allowed us to calculate the heat fluxes from the heated solid to the superfluid helium. The equation for the
heat flow, owing to the elastic process, is the same as the result \cite{Khal3} obtained using the methods of classical acoustics.

It has been shown that all of the exit phonons angles in inelastic processes are allowed in a liquid helium,
which was observed in experiments  \cite{Wyatt10, Wyatt11, Wyatt12, Wyatt13, Wyatt14}. According to \eqref{eq23},
the first inelastic process gives the main contribution to the background channel.
The heat flow \eqref{eq15} from the solid heated to $5$~K to the cold superfluid helium due to the first inelastic
process is $2.3$ times greater than the heat flow produced by the elastic process.
This flow decreases as $T^4$ when the temperature is lowered. Namely, when the temperature of a solid increases,
the contribution of the background radiation increases to the contribution of the elastic process, which corresponds
to the behavior observed in experiments  \cite{Wyatt10, Wyatt11, Wyatt12, Wyatt13, Wyatt14, Wyatt16}.
The absolute values for the heat flux owing to the first inelastic process could only partially explain the big values of
the background radiation, which was observed in experiments  \cite{Wyatt10, Wyatt11, Wyatt12, Wyatt13, Wyatt14}
(see figure \ref{fig1}). Calculated in \cite{ilt_prev} the contribution to the heat transfer coefficient owing
to inelastic processes has also proved to be relatively small and could not explain the large values of the heat
transfer coefficient, which was observed in the experiments on the Kapitza gap.

Thus, an inelastic process has only partially justified the expectations, and a new investigation of the heat transfer between
solid and superfluid helium will be needed to reconcile the theory with the experiments.

\section*{Acknowledgement}
We thank A. F. G. Wyatt for useful discussions which led to the initiation of this work.

%

\ukrainianpart

\title{Взаємодії фононів на поверхні розділу \\
тверде тіло-надплинний гелій}

\author{І.М. Адаменко, Є.К. Нємченко}

\address{Харківський національний університет ім. В.Н. Каразіна,
пл. Свободи, 4, 61022 Харків, Україна}

\makeukrtitle

\begin{abstract}
У роботі запропоновано новий метод отримання гамільтоніана
 взаємодії фононів на поверхні розділу тверде тіло-надплинний гелій.
 Отримано вирази для збурень гідродинамічних змінних у термінах
 вторинного квантування у випадку, якщо рідина займає напівпростір. На
 основі отриманого гамільтоніана обчислено вклади від усіх процесів у
 потік тепла з нагрітого твердого тіла у надплинний гелій. Отримано
 кутові розподіли випромінених твердим тілом фононів у різних процесах.
 Показано, що у випадку непружних процесів немає заборон на кути
 вильоту фононів у надплинний гелій. Отримані результати порівнюються з
 результатами експериментальних та теоретичних робіт.

\keywords фонони, кутовий розподіл, потік тепла, стрибок Капиці, поверхня розділу

\end{abstract}

\end{document}